# Improvement of xenon purification system using a combination of a pulse tube refrigerator and a coaxial heat exchanger

Chen W.-T.[b], Briend P.[a], Chbib D.[b], Cussonneau J.-P.[b], Donnard J.[b], Duval S.[b], Haruyama T.[c], Lemaire O.[b], Le Calloch M.[b], Le Ray P.[b], Mihara S.[c], Mohamad-Hadi A.-F.[b], Morteau E.[b], Oger T.[b], Scotto-Lavina L.[b], Stutzmann J.-S.[b], Tauchi T.[c], Thers D.[b]

[a] AIR LIQUIDE Advanced Technologies Division, 2 rue de Clémencière, F-38360 Sassenage, France
[b] Laboratoire SUBATECH, 4 rue Alfred Kastler, 44307 Nantes cedex 3, France
[c] High Energy Accelerator Research Organization, KEK, 1-1 Oho, Tsukuba, Ibaraki 305-0801 Japan

We have developed a compact cryogenic system with a pulse tube refrigerator and a coaxial heat exchanger. This liquefaction-purification system not only saves the cooling power used to reach high gaseous recirculation rate, but also reduces the impurity level with high speed. The heat exchanger operates with an efficiency of 99%, which indicates the possibility for fast xenon gas recirculation in a high-pressurized large-scale xenon storage with much less thermal losses.

INTRODUCTION

Liquid xenon (LXe) is an attractive material as radiation detector media due to its high atomic number (54) and high density (3 g/cm$^3$), which makes it very efficient to stop penetrating radiation [1]. In the recent 15 years, the size of LXe detectors grows up rapidly because the development of xenon technology makes giant advances. MEG experiment [2] is the milestone in the history of large-size LXe facility because it not only uses 2.7 tons of LXe, but also develops of the powerful cryocoolers [3] and innovative storage system [4], which are specifically designed for LXe temperatures.

For a future ton-scale LXe detector for dark matter detection or medical imaging using both scintillation light and ionization signals, such as XENON1T (~ 3 tons )or DARWIN [5] (~8 tons), the requirement of LXe purity is much more rigorous (~0.1 ppb) compared to MEG (~100 ppb)[*]. The traditional way to purify the LXe is to make a recirculation loop to continuously draw up small amount of LXe from the detector, and pass the xenon through a gas purifier after evaporated, then finally send it back to the detector after recondensed.  The purity of LXe in the detector depends on two variables. One is the outgassing of impurities ($O_2$, $CO_2$, CO, $H_2O$ …) from the surfaces in contact with LXe, the other is the efficiency of the purification system, which is mainly related with the recirculation rate. If the recirculation rate is high enough, the impurities from the surface outgassing can be reduced sufficiently. To reach high recirculation rate with limited cooling power, using a heat exchanger is a good way to recuperate more than 95% of heat for evaporating LXe to recondense the purified xenon [6].

For a ton-scale detector, the time required to obtain decent LXe purity with this type of recirculation process is in the order of a year. Another issue comes out with the large amount of (expensive) xenon: how to store it in case of maintenance or low-time power failures? MEG has developed a double-wall insulated LXe storage system equipped with a pulse tube refrigerator (PTR) [4]. This storage not only be able to store 900 L LXe for more than 44 hours without electricity (~20W heat loss), but also allows filling or recuperating the experiment in liquid phase by using pressure difference and a centrifugal pump. For long-term shut-down, MEG also prepares a gaseous xenon (GXe) storage matrix which can recover the full amount of the xenon by solidifying it with liquid nitrogen ($LN_2$).

---

[*] In order to reach substantially long attenuation length for both scintillation light and ionization signals.

Based on the success of MEG, XENON1T is preparing a storage system. However, due to the lack of space, rigorous requirements of LXe quality and safety, neither a GXe storage matrix nor a centrifugal pump are suitable to be used in the underground facility. Then, as far as we know, the only possible technique to fill or recuperate xenon in liquid phase is the use of pressure difference. It requires enough cooling power to recondense the evaporating xenon during LXe transfer in order to keep the pressure balance. In view of saving LXe preparation time the storage vessel has to be connected with a recirculatiion-purification system. To satisfy these requirements, Subatech and AirLiquide have developed a compact xenon storage system – ReStoX (Recuperation and Storage of Xenon)[*]. ReStoX is composed of a high-pressurized storage which can store 3 tons of xenon in 65 atm, and a built-in $LN_2$ coldhead which can provide 1 kW net cooling power for fast LXe recuperation. In order to reduce $LN_2$ consumption for the purification, ReStoX also equips a coaxial heat exchanger, which can stand 65 atm and is easily to be cleaned compared with a commercial plate heat exchanger.

In this work, we are going to verify the performance of the coaxial heat exchanger in order to complete the design of ReStoX.

APPARATUS AND PROCEDURE

The XEnon Medical Imaging System (XEMIS) [7], which is a LXe time-projection chamber (TPC) being developed for the purpose of LXe medical imaging R&D, is used to verify the design of ReStoX. The impurity level (oxygen equivalent) can be estimated from the attenuation length of ionization signal measured with the TPC [8].

Cooling and heat exchanger
A PTR developed and optimized specifically for LXe temperature is used to provide stable cooling power up to 200W at -108ºC [7]. A built-in heater located between PTR and copper coldhead is used to compensate the extra cooling power of the PTR. The overall cooling power can be controlled by setting the temperature of coldhead, and the heating power can be automatically adjusted by a PID controller. The distance between the cryostat and the PTR is extended to ~2 m to reduce mechanical noise. The PTR is assembled on the top of coaxial heat exchanger in order to not only save space, but also to insulate the PTR by the heat exchanger (Figure 1, left). Multilayer insulation (MLI) is surrounding the PTR in order to reduce the heat loss due to thermal radiation.

The stainless coaxial heat exchanger is made by bending a 22.5m long coaxial tube to a ~1m long hollow cylindrical shape. The inner diameters of tube side and shell side are 17.08mm and 48.26mm, respectively. It is designed in order to have 95% of efficiency for up to 50 NL/min (4.92 g/s) of LXe/GXe recirculation, with an assumption of 0.5 bar pressure difference between tube side and shell side, and 6K temperature difference at cold end[†]. The vertical distance between the heat exchanger and cryostat is ~50cm. Several temperature sensors and pressure sensors are pasted on the inlet/outlet of tube/shell side in order to estimate the efficiency and the pressure drop of heat exchanger.

Schematic
The scheme of the whole cryogenic system is presented in Figure 1 (right). Xenon from the storage bottle is liquefied and filled into the cryostat by the remote PTR after being purified by the getter (rare gas purifier) and passing through the shell side of heat exchanger. A level meter installed inside the cryostat allows knowing the level of LXe and estimating its mass during the filling and recirculation. The filling process is stopped when ~36.5 kg of LXe is filled into the cryostat. The recirculation is driven by an oil-free membrane pump which can provide a maximum delivery of 50 NL/min [9]. Two rare-gas purifiers [10] connected in parallel are used to purify GXe with less pressure drop. With the membrane pump, LXe can be pumped into the tube side of heat exchanger and evaporated by the heat from the shell side; after passing through the purifier, warm xenon is sent back to the shell side and recondensed into the cryostat. The circulation flow is controlled by the valve between the outlet of the tube side and the pump. In case of emergency, xenon can be recuperated into a 4m³ rescue tank.

---

[*] Is also presented in ICEC24-ICMC2012
[†] The inlet of tube side and the outlet of shell side, which are close to LXe temperature

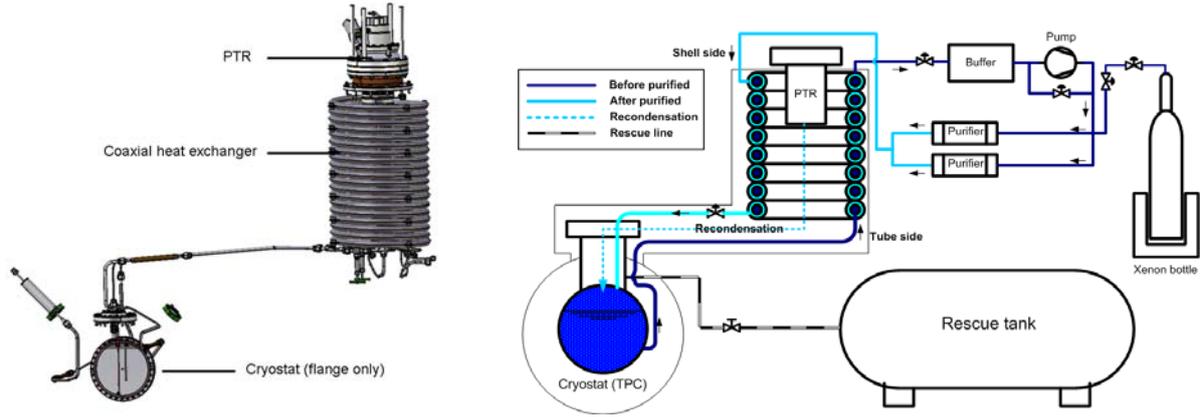

Figure 1: (left) 3D drawing of the PTR, heat exchanger and the connection with cryostat. (right) Schematic of XEMIS cryogenic system.

PERFORMANCE OF HEAT EXCHANGER

To estimate the performance of the present heat exchanger as a function of recirculation rate, it is necessary to measure the pressure and temperature at inlet and outlet. Concerning the pressure drop, the one in the shell side is measured from the difference of pressure at inlet and at cryostat (outlet). For the tube side, because the valve between the outlet and the pressure sensor is used to adjust the flow, the pressure at the outlet can only be measured directly at highest flow (~33 NL/min, the valve is fully opened). In order to estimate the pressure at outlet, we assume that the pressure drop is proportional to the xenon mass inside the heat exchanger, which can be obtained from the LXe level measurement. The result of pressure drop is shown in Figure 2 (left). The pressure drop in the tube side is much bigger due to the vertical distance between cryostat and heat exchanger (~50 cm). The measurement of temperature difference is shown in Figure 2 (right). The temperature at cold end is quite stable (tube side: ~-103 ºC, shell side: ~ -101 ºC), which indicates that there is heat exchange between LXe and GXe inside heat exchanger even at small flow (~2.5 NL/min). Compare with the assumption of pressure difference between tube/shell side and temperature difference at the cold end for the design of heat exchanger (0.5 bar, 6 K), the observed values are much smaller (~0.15 bar, 2 K). The efficiency of heat exchanger ($\varepsilon$) as a function of recirculation rate can be estimated as:

$$\varepsilon = 1 - \frac{C_p \times \Delta T_{warm} \times F}{Q} \qquad (1)$$

Where Q (W) is the thermal duty to recondense a given gaseous xenon flow F (g/s)[*]. $C_p$ (J/g/K) and $\Delta T_{warm}$ (K) are the specific heat of xenon and the temperature difference at warm end[†], respectively. The measurement of cooling power is shown in Figure 3 (left), which shows that the change of cooling power is nearly unobservable at high flow. The efficiency estimated by Equation 1 is shown in Figure 3 (right). It is 99% even at high flow, which is even better than its design value (95%) or the performance of the plate heat exchanger described in [6].

CONCLUSION

The measurement indicates that the performance of coaxial heat exchanger is better than its designed value. Based on its good performance and high strength, it may not only be a good solution for ReStoX, but also be a good device for all LXe experiments like DARWIN [5] project.

---

[*] Is converted from the flow (NL/min) measured by flow meter
[†] The outlet of tube side and the inlet of shell side, which are close to room temperature

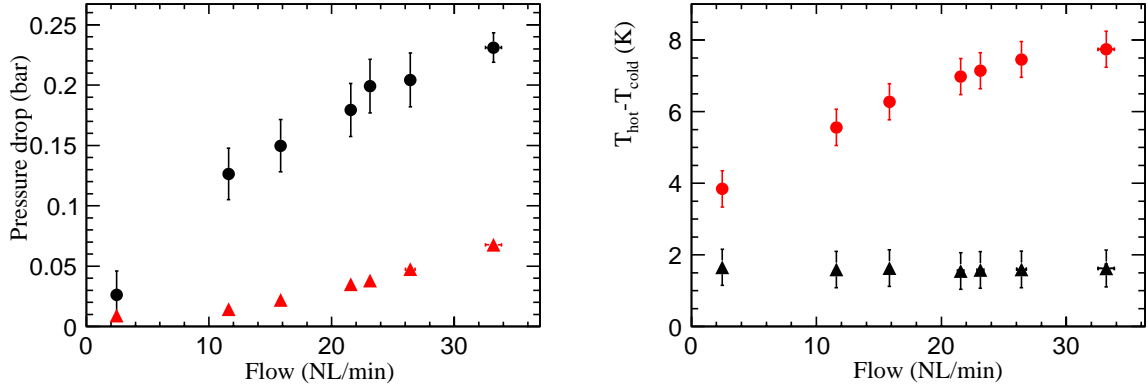

Figure 2: (left) the pressure drop in the tube side (black solid circle) and shell side (red triangle) as a function of flow. The pressure drop of shell side is calculated from the difference of pressure at the inlet of shell side and at the cryostat. For the tube side, only the point at ~33L/min is measured from the pressure sensor. The others are estimated from the measurement of LXe mass inside heat exchanger. (right) The temperature difference of the cold end (black triangle) and hot end (red solid circle).

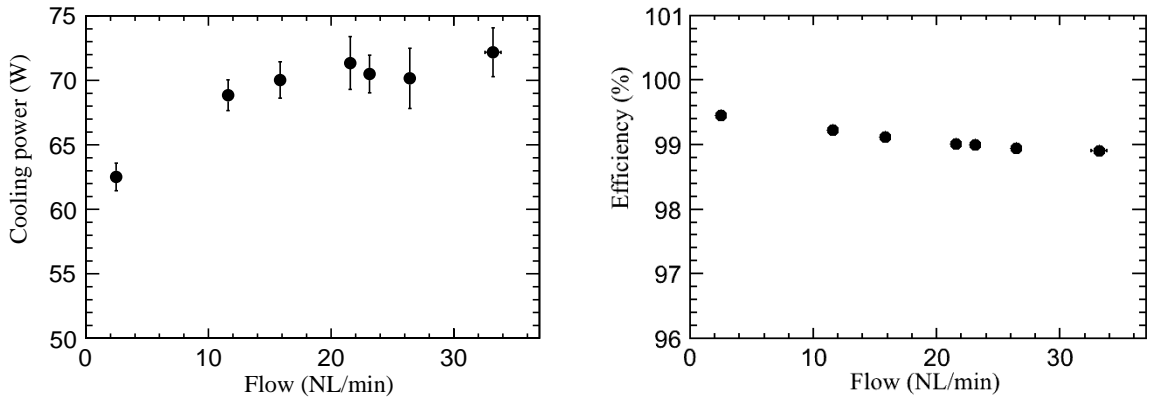

Figure 3: (left) The cooling power (estimated) in order to keep the cryostat at stable pressure (~1.4 bar) as a function of flow. (right) The efficiency (%) of heat exchanger in different flow, which is estimated using Equation 1.

## ACKNOWLEDGMENT

We thank Prof. Elena Aprile and Dr. Ranny Budnik for the help of experimental device. This work is supported by the region of Pays de la Loire, France.